\documentclass[12pt]{extarticle}
\usepackage{setspace}

\usepackage{setspace}
\usepackage[T1]{fontenc}
\usepackage{times}
\usepackage{palatino} 
\usepackage{lmodern}
\usepackage[latin1]{inputenc}
\usepackage{epsfig}
\usepackage[english]{babel}
\usepackage{color}
\usepackage{graphicx}
\usepackage{dcolumn}
\usepackage{moreverb}
\usepackage{amsmath,amssymb,amsfonts}
\usepackage[all]{xy}

\usepackage[svgnames]{xcolor}
\usepackage{tikz}

\def\nudge{.5}

\tikzset{axis/.style={ultra thick, Red!75!black, -latex, shorten <=-\nudge cm, shorten >=-2*\nudge cm}}
\tikzset{line/.style={thick,Green}}

\begin{document}
\numberwithin{equation}{section}
\newcommand{\boxedeqn}[1]{%
  \[\fbox{%
      \addtolength{\linewidth}{-2\fboxsep}%
      \addtolength{\linewidth}{-2\fboxrule}%
      \begin{minipage}{\linewidth}%
      \begin{equation}#1\end{equation}%
      \end{minipage}%
    }\]%
}


\newsavebox{\fmbox}
\newenvironment{fmpage}[1]
     {\begin{lrbox}{\fmbox}\begin{minipage}{#1}}
     {\end{minipage}\end{lrbox}\fbox{\usebox{\fmbox}}}

\raggedbottom
\onecolumn

\parindent 8pt
\parskip 10pt
\baselineskip 16pt
\noindent\title*{{\LARGE{\textbf{Recurrence approach and higher rank cubic algebras for the $N$-dimensional superintegrable systems}}}}
\newline
\newline
\newline
Md Fazlul Hoque, Ian Marquette and Yao-Zhong Zhang
\newline
School of Mathematics and Physics, The University of Queensland, Brisbane, QLD 4072, Australia
\newline
E-mail: m.hoque@uq.edu.au; i.marquette@uq.edu.au; yzz@maths.uq.edu.au
\newline
\newline
\begin{abstract}
By applying the recurrence approach and coupling constant metamorphosis, we construct higher order integrals of motion for the Stackel equivalents of the $N$-dimensional superintegrable Kepler-Coulomb model with non-central terms and the double singular oscillators of type ($n, N-n$). We show how the integrals of motion generate higher rank cubic algebra $C(3)\oplus L_1\oplus L_2$ with structure constants involving Casimir operators of the Lie algebras $L_1$ and $L_2$. The realizations of the cubic algebras in terms of deformed oscillators enable us to construct finite dimensional unitary representations and derive the degenerate energy spectra of the corresponding superintegrable systems.
\end{abstract}

\section{Introduction}
A systematic approach for obtaining algebraic derivations of spectra of two dimensional superintegrable systems with quadratic and cubic algebras involving three generators was introduced in \cite{Das1, Mar1, Mar2}. This algebraic method is based on Casimir operators and realizations in terms of deformed oscillator algebras \cite{Das2} for determining finite dimensional unitary representations (unirreps) \cite{Mar1,Mar3}. This approach was extended to classes of higher order polynomial algebras with three generators \cite{Isa1}. However, superintegrable systems in higher dimensional spaces are usually associated with symmetry algebras taking the form of higher rank polynomial algebras which have typically a quite complicated embedded structure. It was discovered recently that there are systems which are related to classes of quadratic algebras that display a decomposition as direct sums of Lie algebras and polynomial algebras of three generators only \cite{FH1,FH2}. The structure constants contain Casimir operators of certain Lie algebras \cite{FH1,FH2}. This specific structure was exploited to obtain algebraic derivations of the spectra. However, in general it is quite complicated to apply this direct approach to obtain the corresponding polynomial algebras, the Casimir operators and their realizations in terms of deformed oscillators. In fact, it is not even guaranteed that the integrals close to certain polynomial algebra. 

The difficulties of this direct approach can be overcome using a constructive approach and in particular using ladder operators in order to build integrals of motion for models allowing separation of variables in Cartesian coordinates. Such ideas have been used by several authors in the case of first or second order ladder operators \cite{Jau1, Fri1, Boy1, Eva1,Ver1, Mar4}. It facilitates the construction of the corresponding polynomial algebras. One has to distinguish the case of constructing integrals using higher order (i.e. greater than 2) ladder operators as such operators themselves need to be generated using various methods. One of them consists in exploiting supersymmetric quantum mechanics (SUSYQM) \cite{Jun1} and using combinations of ladders and supercharges \cite{Dem1, Mar5, Rag1, Que1} or combining only supercharges \cite{Mar4, Kre1, Adl1, Mar6, Que2}.

In recent years, Kalnins, Kress and Miller\cite{Mil1} introduced a recurrence relation approach and applied it to models separable in polar coordinates. They also pointed out the close relation between such approach and the study of special functions and orthogonal polynomials. Calzada, Kuru and Negro \cite{Cal1, Cal2} introduced an operator version of the recurrence relations. In this scheme an intermediate set of non-polynomial integrals of motion was obtained. From these formal algebraic relations, and by decomposing these integrals into polynomial and non-polynomial parts, a final well-defined set of integrals and the corresponding polynomial algebra can be obtained. It has also been demonstrated how SUSYQM can be combined with these ideas to generate extended Lissajous models related to Jacobi exceptional orthogonal polynomials \cite{Que3}. 

All these new approaches have previously been restricted mainly to two and three dimensional superintegrable systems. The purpose of this paper is to extend the recurrence approach for a class of higher-dimensional systems and construct higher rank polynomial algebras. We consider the following two models \cite{FH1,FH2}
\begin{eqnarray}
&&{\bf Model\quad 1}:\quad H_{dso}=\frac{p^2}{2}+\frac{\omega^2 r^2}{2}+\frac{c_1}{x^2_1+...+x^2_n}+\frac{c_2}{x^2_{n+1}+...+x^2_N},\label{hamil1}
\\&&
{\bf Model\quad 2}:\quad H_{KC}=\frac{1}{2}p^{2}-\frac{c_{0}}{r}+\frac{c_{1}}{r(r+x_{N})}+\frac{c_{2}}{r(r-x_{N})}, \label{hamil2}
\end{eqnarray}
where $ \vec{r}=(x_{1},x_{2},...,x_{N})$, $\vec{p}=(p_{1},p_{2},...,p_{N})$, $r^{2}=\sum_{i=1}^{N}x_{i}^{2}$, $p_{i}=-i \hbar \partial_{i}$ and $c_0$, $c_1$, $c_2$ are positive real constants. The model (\ref{hamil1}) is a family of $N$-dimensional superintegrable double singular oscillators and model (\ref{hamil2}) is $N$-dimensional superintegrable Kepler-Coulomb system with non-central terms in $N$-dimensional Euclidean space. In two recent papers, using a direct approach and ansatz to construct the integrals and the quadratic algebra, an algebraic derivation of the energy spectra have been presented \cite{FH1,FH2}. In this paper, we will show using coupling constant metamorphosis \cite{Mill3} that a constructive approach can be developed for these N-dimensional models and be used to simplify the calculation and analysis.

The plan of the paper is as follows. In section 2, higher order integrals are constructed from ladder operators using separated eigenfunctions of the system (\ref{hamil1}) and the corresponding higher rank cubic algebra is presented. In section 3, higher order integrals and the higher rank cubic algebras are constructed for the Stackel equivalent system of (\ref{hamil2}) from coupling constant metamorphosis and ladder operators. The realizations in terms of deformed oscillators are obtained and applied to compute energy spectra of the two systems algebraically. Finally, in section 4, we provide some discussions on the results of this paper and some open problems.

\section{Recurrence approach to $H_{dso}$}
In this section, we develop for the model (\ref{hamil1}) the recurrence relations in order to generate higher order integrals of motion and the corresponding higher rank polynomial algebra. We show how in fact the integrals close into a cubic algebra with structure constants involving Casimir operators of certain Lie algebras and derive the energy spectrum of the system algebraically. This provides also a proof for the superintegrability of this $N$-dimensional system.

\subsection{Separation of variables}
We recall \cite{FH2} in this subsection the separable solutions of (\ref{hamil1}) in double hyperspherical coordinates using the sum of two singular oscillators $H_{dso}=H_1+H_2$ of dimensions $n$ and $N-n$ respectively, where
\begin{eqnarray}
 &&H_1=\frac{1}{2}(p^2_1+...+p^2_n)+\frac{\omega^2}{2}r^2_1+\frac{c_1}{r^2_1},
 \\&&
 H_2=\frac{1}{2}(p^2_{n+1}+...+p^2_N)+\frac{\omega^2}{2}r^2_2+\frac{c_2}{r^2_2}.
 \end{eqnarray}
The Schrodinger equation of $H_1$ in $n$-dimensional hyperspherical coordinates is 
\begin{eqnarray}
-\frac{1}{2}\left[\frac{\partial^2}{\partial r^2_1}+\frac{n-1}{r_1}\frac{\partial}{\partial r_1}-\frac{1}{r^2_1}\Lambda^2(n)-\omega'^2 r^2_1-\frac{2c'_1}{r^2_1}\right]\psi_1(r_1,\Omega_{n-1})=E'_1\psi_1(r_1,\Omega_{n-1}),\label{kpH1}
\end{eqnarray}
where $c'_1=\frac{c_1}{\hbar^2}$, $\omega'=\frac{\omega}{\hbar}$,  $E'_1=\frac{E_1}{\hbar^2}$ and $\Lambda^2(n)$ is the grand angular momentum operator. The wave function $\psi_1(r_1,\Omega_{n-1})$ is proportional to
\begin{eqnarray}
e^{-\frac{\omega' r^2_1}{2}}r_1^{\alpha_1+\frac{n}{2}}L^{\alpha_1}_{n_1}(\omega' r^2_1)y_1(\Omega_{n-1}),\label{kpWf}
\end{eqnarray}
where $L^{\alpha}_n(x)$ is the $n$-th order Laguerre polynomial \cite{Ask1}, 
$\alpha_1=2\delta_1+l_{n}+\frac{n-2}{2}$, 
$\delta_1=\left\{\sqrt{(\frac{1}{2}l_{n}+\frac{n-2}{4})^2+\frac{1}{2}c'_1}-\frac{n-2}{4}\right\}-\frac{1}{2}l_{n}$ and $n_1=\frac{E'_1}{2\omega'}-\left(\delta_1+\frac{1}{2}l_{n}+\frac{n}{4}\right)$. The wave function for $H_2$ has similar form. The energy spectrum of Hamiltonian (\ref{hamil1}) is
\begin{eqnarray}
E_{dso}=2\hbar \omega\left(p+1+\frac{\alpha_1+\alpha_2}{2}\right),
\end{eqnarray}
where the parameters $\alpha_2=2\delta_2+l_{N-n}+\frac{N-n-2}{2}$, $\delta_2=\left\{\sqrt{(\frac{1}{2}l_{N-n}+\frac{N-n-2}{4})^2+\frac{1}{2}c'_2}-\frac{N-n-2}{4}\right\}-\frac{1}{2}l_{N-n}$, $c'_2=\frac{c_2}{\hbar^2}$ and $p=n_1+n_2$, $n_2=\frac{E'_2}{2\omega'}-\left(\delta_2+\frac{1}{2}l_{N-n}+\frac{N-n}{4}\right)$, $E'_2=\frac{E_2}{\hbar^2}$.

\subsection{Recurrence formulas and algebra structure}
 
Consider the gauge transformations to $H_i$
 \begin{eqnarray}
 \tilde{H_i}=\mu^{-1}_i H_i\mu_i
 \end{eqnarray}
 with $\mu_1=r^{\frac{1-n}{2}}_1$, $\mu_2=r^{\frac{1-N+n}{2}}_2$. We obtain the following gauge equivalent operators that possesses the same eigenvalues
\begin{eqnarray}
\tilde{H_i}=-\frac{1}{2}\left\{\partial^2_{r_i}-\omega'^2 r^2_i-\frac{\beta_i}{r^2_i}\right\},\quad i=1,2.
\end{eqnarray}
Here $\beta_1=\frac{(n-1)(n-3)}{4}+2c'_1+\frac{J_{(2)}}{\hbar^2}$ and $\beta_2=\frac{(N-n-1)(N-n-3)}{4}+2c'_2+\frac{K_{(2)}}{\hbar^2}$ with
\begin{eqnarray}
&&J_{(2)}=\sum_{i<j}J^2_{ij},\quad J_{ij}=x_i p_j-x_j p_i, i, j= 1,2,....,n,
\\&&
 K_{(2)}=\sum_{i<j}K^2_{ij},\quad K_{ij}=x_i p_j-x_j p_i, i, j= n+1,....,N,
 \end{eqnarray}
 $J_{(2)}$ and $ K_{(2)}$ are related to $\Lambda^2(n)$ and $\Lambda^2(N-n)$ respectively. 

Let $Z=X_{n_1}X_{n_2}y_1(\Omega_{n-1})y_2(\Omega_{N-n-1})$ with
\begin{eqnarray}
&&X_{n_i}=e^{-\frac{\omega' r^2_i}{2}}r_i^{\alpha_i+\frac{1}{2}}L^{\alpha_i}_{n_i}(\omega' r^2_i), \quad i=1,2.
\end{eqnarray}
Then the wave functions of the gauge transformed $\tilde{H_i}$ are given simply by $Z=Z_1Z_2$, $Z_i= X_{n_i}y_i=\mu_i^{-1}\psi_i(r_i,\Omega), i=1, 2$. Alternatively, we can gauge transform $\tilde{H_i}$ back to get the initial Hamiltonian
\begin{eqnarray}
 H_{dso}=\mu_1\mu_2\tilde{H} \mu^{-1}_1\mu^{-1}_2,\quad \tilde{H}=\tilde{H_1}+\tilde{H_2}.
\end{eqnarray}
Thus we have the eigenvalue equations $\tilde{H_i} Z_i=\lambda_{r_i}Z_i=\tilde{E'_i} Z_i$ with
\begin{eqnarray}
\lambda_{r_i}=\omega'(2n_i+\alpha_i+1),\quad i=1,2.
\end{eqnarray} 
Hence the energy eigenvalues of $\tilde{H}Z=\tilde{E'}Z$ are given
\begin{eqnarray}
\tilde{E'}=\{2+2(n_1+n_2)+\alpha_1+\alpha_2\}\omega'.
\end{eqnarray}
Now we define the ladder operators
\begin{eqnarray}
\tilde{ D^\pm_i}(\omega', r_i)=-2\tilde{H_i}\mp 2\omega' r_i\partial_{r_i}+2\omega'^2 r^2_i\mp\omega',\quad i=1,2.
\end{eqnarray}
The action of the symmetry operators on the wave functions $Z_i=X_{n_i}y_i, i=1,2$ provides the following recurrence formulas 
\begin{eqnarray}
&&\tilde{D^+_i}(\omega', r_i)X_{n_i}y_i=-4\omega'(n_i+1)X_{n_i+1}y_i,\quad i=1,2,
\\&&
\tilde{D^-_i}(\omega', r_i)X_{n_i}y_i=-4\omega'(n_i+\alpha_i)X_{n_i-1}y_i,\quad i=1,2.
\end{eqnarray}
The following diagram indicates how $\tilde{D^\pm_i}$ change the quantum numbers of the wave functions. 

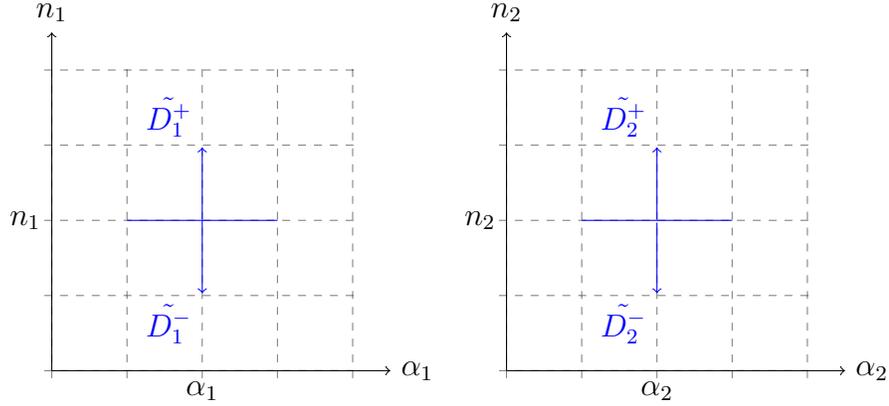
\begin{tikzpicture}
\draw[dashed,color=gray] (-0.1,-0.1) grid (4.1,4.1);

    \draw[->] (0,0) -- (0,4.5) node[above] {$n_1$};
    \draw[ ->] (0,0) -- (4.5,0) node[right] {$\alpha_1$};
       
    \draw[-> ,blue] (2,2.01) -- (2,2.97)node[above left] {$\tilde{D^+_1}$};
    \draw[-> ,blue] (2,2) --(2,1.03) node[below left] {$\tilde{D^-_1}$};
    
    \draw[- ,blue] (1,2) -- (3,2);
        \node at (2,0)[below]{$\alpha_1$};
    \node at (0,2)[left]{$n_1$};
      
\end{tikzpicture}
\begin{tikzpicture}
\draw[dashed,color=gray] (-0.1,-0.1) grid (4.1,4.1);

    \draw[->] (0,0) -- (0,4.5) node[above] {$n_2$};
    \draw[ ->] (0,0) -- (4.5,0) node[right] {$\alpha_2$};
       
    \draw[-> ,blue] (2,2.01) -- (2,2.97)node[above left] {$\tilde{D^+_2}$};
    \draw[-> ,blue] (2,1.97) --(2,1.03) node[below left] {$\tilde{D^-_2}$};
    
    \draw[- ,blue] (1,2) -- (3,2);
        \node at (2,0)[below]{$\alpha_2$};
    \node at (0,2)[left]{$n_2$};
      
\end{tikzpicture}

Let us consider the following suitable combinations of the operators $\tilde{L_1}=\tilde{D^+_1}\tilde{ D^-_2}$, $\tilde{L_2}=\tilde{D^-_1}\tilde{ D^+_2}$, $\tilde{H}=\tilde{H_1}+\tilde{H_2}$ and $\tilde{B}=\tilde{H_1}-\tilde{H_2}$. The action of the operators on the wave functions are given by 
\begin{eqnarray}
&&\tilde{L_1}Z=16\omega'^2(n_1+1)(n_2+\alpha_2)X_{n_1+1}X_{n_2-1}y_1y_2,
\\&&
\tilde{L_2}Z=16\omega'^2(n_2+1)(n_1+\alpha_1)X_{n_1-1}X_{n_2+1}y_1y_2,
\\&&
\tilde{L_1}\tilde{L_2} Z=256\omega'^4 n_1(n_2+1)(n_1+\alpha_1)(n_2+\alpha_2+1)Z,
\\&&
\tilde{L_2}\tilde{L_1} Z=256\omega'^4 n_2(n_1+1)(n_2+\alpha_2)(n_1+\alpha_1+1)Z.
\end{eqnarray}
It follows that in operator they form the cubic algebra $C(3)$,
\begin{eqnarray}
&[\tilde{L_1},\tilde{H}]=0= [\tilde{L_2},\tilde{H}],&\label{kpf1}
\\&[\tilde{L_1},\tilde{B}]=-4\omega'\tilde{L_1} ,\qquad [\tilde{L_2},\tilde{B}]=4\omega'\tilde{L_2},&\label{kpf2}
\end{eqnarray} 
\begin{eqnarray}
&\tilde{L_1}\tilde{L_2}& =\left[(\tilde{B} +\tilde{H}-2\omega')^2-\frac{4\omega'^2}{\hbar^2}\left\{J_{(2)}+2c'_1\hbar^2+\frac{(n-2)^2}{4}\hbar^2\right\}\right] \nonumber\\&& \times\left[(\tilde{B} -\tilde{H}-2\omega')^2-\frac{4\omega'^2}{\hbar^2}\left\{K_{(2)}+2c'_2\hbar^2+\frac{(N-n-2)^2}{4}\hbar^2\right\}\right],\nonumber\\&&\label{kpf5}
\\&
\tilde{L_2}\tilde{L_1}&=\left[(\tilde{B} +\tilde{H}+2\omega')^2-\frac{4\omega'^2}{\hbar^2}\left\{J_{(2)}+2c'_1\hbar^2+\frac{(n-2)^2}{4}\hbar^2\right\}\right] \nonumber\\&& \times\left[(\tilde{B} -\tilde{H}+2\omega')^2-\frac{4\omega'^2}{\hbar^2}\left\{K_{(2)}+2c'_2\hbar^2+\frac{(N-n-2)^2}{4}\hbar^2\right\}\right].\nonumber\\&&\label{kpf6}
\end{eqnarray}

The first order integrals of motion $J_{ij}$ and $K_{ij}$ generate algebras isomorphic to the $so(n)$  and $so(N-n)$ Lie algebras respectively. Namely, 
\begin{eqnarray}
&[J_{ij},J_{kl}]&= i(\delta_{ik}J_{jl}+ \delta_{jl}J_{ik}-\delta_{il}J_{jk}-\delta_{jk}J_{il})\hbar, 
\end{eqnarray}
where $i, j, k, l=1, ..., n$ and 
\begin{eqnarray}
&[K_{ij},K_{kl}]&=i( \delta_{ik}K_{jl}+ \delta_{jl}K_{ik}-\delta_{il}K_{jk}-\delta_{jk}K_{il})\hbar, 
\end{eqnarray}
with $i, j, k, l=n+1, ..., N-n$. Moreover,
\begin{eqnarray}
&[J_{ij},\tilde{L_1}]=0=[J_{ij},\tilde{L_2}],\quad[K_{ij},\tilde{L_1}]=0=[K_{ij},\tilde{L_2}],&
\\&
[J_{ij},\tilde{H}]=0=[K_{ij},\tilde{H}],\quad [J_{ij},\tilde{B}]=0=[K_{ij},\tilde{B}].&
\end{eqnarray}
So the full symmetry algebra is a direct sum of the cubic algebra $C(3)$, $so(n)$ and $so(N-n)$ Lie algebras. Thus the $su(N)$ Lie algebra generated by the integrals of the $N$-dimensional isotropic harmonic oscillators is deformed into the higher rank cubic algebra $C(3) \oplus so(n) \oplus so(N-n)$ for Hamiltonian (\ref{hamil1}).

The recurrence method gives rise to higher order integrals and higher rank polynomial algebra, more specifically 4th-order integrals and a cubic algebra $C(3)\oplus so(n)\oplus so(N-n)$ involving Casimir operators of $so(n)$ and $so(N-n)$ Lie algebras. In our recent work \cite{FH2} we showed that the second order integrals $A$ and $B$ generate the quadratic algebra $Q(3)\oplus so(n)\oplus so(N-n)$ with commutation relations $[A,B]=C$, $[A,C]=f_1(A, B, H, J_{(2)},  K_{(2)})$, $[B,C]=f_2(A, B, H, J_{(2)}, K_{(2)})$, where $f_{1}$ and $f_{2}$ are quadratic polynomials in the generators $A$ and $B$ and the central elements $H$, $J_{(2)}$ and $K_{(2)}$. In order to derive the spectrum using the quadratic algebra $Q(3)$, realizations of $Q(3)$ in terms of deformed oscillator algebras \cite{Das1,Das2} $\{\aleph, b^{\dagger}, b\}$ of the form
\begin{eqnarray}
[\aleph,b^{\dagger}]=b^{\dagger},\quad [\aleph,b]=-b,\quad bb^{\dagger}=\Phi (\aleph+1),\quad b^{\dagger} b=\Phi(\aleph),\label{kpfh}
\end{eqnarray}
have been used. Where $\aleph $ is the number operator and $\Phi(x)$ is well behaved real function satisfying the constraints
\begin{eqnarray}
\Phi(0)=0, \quad \Phi(x)>0, \quad \forall x>0.\label{kpbc}
\end{eqnarray}
It is non-trivial to obtain such a realization and find the structure function $\Phi(x)$. However in the recurrence approach presented in this paper, the cubic algebra $C(3)$ relations (\ref{kpf1}) - (\ref{kpf6}) already have, in fact, the form of a deformed oscillator algebra (\ref{kpfh}).

\subsection{Unirreps and energy spectrum}
We write the cubic algebra relations (\ref{kpf1})-(\ref{kpf6}) in the form of deformed oscillator (\ref{kpfh}) by letting $\aleph=\frac{\tilde{B}}{4w'}$, $b^{\dagger}=\tilde{L_1}$ and $b=\tilde{L_2}$. We then readily obtain the structure function
\begin{eqnarray}
&\Phi(x,u,\tilde{H})&=\left[4\omega'(x+u)+\tilde{H}-2(1-\alpha_1)\omega'\right]\left[4\omega'(x+u)+\tilde{H}-2(1+\alpha_1)\omega'\right]\nonumber\\&&\times\left[4\omega'(x+u)-\tilde{H}-2(1-\alpha_2)\omega'\right]\left[4\omega'(x+u)-\tilde{H}-2(1+\alpha_2)\omega'\right],\nonumber\\&&\label{un1}
\end{eqnarray}
where $u$ is arbitrary constant. In order to obtain the (p+1)-dimensional unirreps, we should impose the following constraints on the structure function 
\begin{equation}
\Phi(p+1; u,\tilde{E'})=0,\quad \Phi(0;u,\tilde{E'})=0,\quad \Phi(x)>0,\quad \forall x>0,\label{pro2}
\end{equation}
where $p$ is a positive integer. The solutions give the energy $\tilde{E'}$ and the arbitrary constant $u$. Let $\varepsilon_1=\pm 1$, $\varepsilon_2=\pm 1$. We have
\begin{eqnarray}
u=\frac{-\tilde{E'}+2\omega'(1+\varepsilon_1\alpha_1)}{4\omega'},\quad\tilde{E'}=(2+2p+\varepsilon_1\alpha_1+\varepsilon_2\alpha_2)\omega',
\end{eqnarray}
\begin{eqnarray}
\Phi(x) =256x\omega'^4(x+\varepsilon_1\alpha_1)(1+p-x)(1+p-x+\varepsilon_2\alpha_2).
\end{eqnarray}
The physical wave functions involve other quantum numbers and we have in fact degeneracy of $p+1$ only when these other quantum numbers would be fixed. The total number of degeneracies may be calculated by taking into account the further constraints on these quantum numbers. The results have been obtained in the gauge transformed Hamiltonian. However, the relations $D^{\pm}_i =\mu_i \tilde{D^{\pm}_1} \mu^{-1}_i, i=1,2$  provide the integrals of motion $L_1= D^{+}_1 D^{-}_2$ and $L_2= D^{-}_1 D^{+}_2$ of the initial Hamiltonian $H_{dso}$ and the algebraic derivation remain valid as gauge transformations preserve the spectrum.

\section{Recurrence approach to $H_{KC}$}
In this section, we construct recurrence relations to generate higher order integrals and higher rank polynomial algebra from ladder operators and coupling constant metamorphosis for the Stackel equivalent of model (\ref{hamil2}).
 
\subsection{Saparation of variables}
The Schrodinger equation of the system (\ref{hamil2}) in the hyperparabolic coordinates reads 
\begin{eqnarray}
&H\psi(\xi, \eta, \Omega_{N-1})&=\left[-\frac{2}{\xi+\eta}\left[\Delta(\xi)+\Delta(\eta)-\frac{\xi+\eta}{4\xi\eta}\Lambda^2(\Omega_{N-1})\right]-\frac{2\beta_{0}}{\xi+\eta}\right.\nonumber\\&&\left.+\frac{2\beta_{1}}{\xi(\xi+\eta)}+\frac{2\beta_{2}}{\eta(\xi+\eta)}\right]\psi(\xi, \eta, \Omega_{N-1})=\varepsilon \psi(\xi, \eta, \Omega_{N-1}),\nonumber\\&&\label{fk1}
\end{eqnarray} 
where $\Lambda^2(N)$ is the grand angular momentum operator and 
\begin{eqnarray*}
&\Delta(\xi)& = \xi^{-\frac{N-3}{2}}\frac{\partial}{\partial \xi}  \xi^{\frac{N-1}{2}}\frac{\partial}{\partial \xi}, \quad \Delta(\eta) = \eta^{-\frac{N-3}{2}}\frac{\partial}{\partial \eta}  \eta^{\frac{N-1}{2}}\frac{\partial}{\partial \eta},\\& \beta_{0}&=\frac{c_{0}}{\hbar{^2}}, \quad \beta_{1}=\frac{c_{1}}{\hbar{^2}},\quad \beta_{2}=\frac{c_{2}}{\hbar{^2}},\quad \varepsilon=\frac{E}{\hbar^{2}}.
\end{eqnarray*}
We can write the equivalent system of (\ref{fk1}) as 
\begin{eqnarray}
&H'\psi(\xi, \eta, \Omega_{N-1})&=\left[\Delta(\xi)+\Delta(\eta)-\frac{\beta_{1}}{\xi}-\frac{\beta_{2}}{\eta}+\frac{\omega'}{2}(\xi+\eta)-\frac{1}{4\xi}\Lambda^2(\Omega_{N-1})\right.\nonumber\\&&\left.-\frac{1}{4\eta}\Lambda^2(\Omega_{N-1})\right]\psi(\xi, \eta, \Omega_{N-1})=\varepsilon'\psi(\xi, \eta, \Omega_{N-1}).\label{fk2}
\end{eqnarray}
The original energy parameter $\varepsilon$ now plays the role of model parameter (or coupling constant ) $\omega'$ and the model parameter $-\beta_0$ plays the role of energy $\varepsilon'$. This change in the role of the parameters is called coupling constant metamorphosis. Moreover, the Hamiltonian (\ref{fk2}) is related to the one in (\ref{fk1}) by a Stackel transformation and thus the two systems are Stackel equivalent \cite{Post1,Mill3}. 

After the change of variables $\xi=2r^2_1$, $\eta=2r^2_2$, the wave functions in the new variables are proportional to
 \begin{eqnarray} 
e^{-\frac{\sqrt{-\omega'} r^2_i}{2}}r_i^{\alpha'_i+\frac{N-1}{2}}L^{\alpha'_i}_{n_i}(\sqrt{-\omega'} r^2_i)y_i(\Omega_{N-1}),\quad i=1,2,\label{Wf1} 
\end{eqnarray}
where $L^{\alpha}_{n}(x)$ is again the $n$-th order Laguerre polynomial and 
$\alpha'_i=\delta_i+l_{N-2}+\frac{N-3}{2}$, $\delta_{i}=\left\{\sqrt{(I_{N-2}+\frac{N-3}{2})^2+4\beta_{i}}-\frac{N-3}{2}\right\}-I_{N-2}$, $n_{i}=-\frac{1}{2}\left(\delta_{i}+I_{N-2}+\frac{N-1}{2}\right)+\frac{\varepsilon'}{\omega'}$, $i=1, 2$.  The energy spectrum of system (\ref{hamil2}) is
 \begin{equation}
E=\frac{-c^{2}_{0}}{\hbar^{2}\left\{n_{1}+n_{2}+\frac{1}{2}(\delta_{1}+\delta_{2}+2I_{N-2}+N-1)\right\}^{2}}.\label{en2}
\end{equation}

\subsection{Recurrence formula and algebra structure}

Let $H'=H_1'+H_2'$ and perform gauge rotations $\tilde{H_i}=\chi_i^{-1} H'_i \chi_i$, $\chi_i=r^{\frac{2-N}{2}}_i$, $i=1,2$ to get 
\begin{eqnarray}
&\tilde{H_i}&=\frac{1}{4}\left\{\partial^2_{r_i}+2\omega'r^2_i-\frac{\frac{(N-2)(N-4)}{4}+4 \beta_i+\frac{J_{(2)}}{\hbar^2}}{r^2_i}\right\},
\end{eqnarray}
where  
\begin{eqnarray}
J_{(2)}=\sum_{i<j}^{N-1}L_{ij}^{2},\quad L_{ij}=x_{i}p_{j}-x_{j}p_{i},\quad i, j=1,\dots, N-1,
\end{eqnarray}
and $J_{(2)}$ is related to $\Lambda^2(\Omega_{N-1})$.  
Let $Z$ be eigenfunctions of the gauge rotated Hamiltonian $\tilde{H}=\tilde{H_1}+\tilde{H_2}$, $\tilde{H}Z=\tilde{\varepsilon} Z$. Writing $Z=X_{n_1}X_{n_2}y_1(\Omega_{N-1})y_2(\Omega_{N-1})=\chi^{-1}_i\psi(r_1, r_2, \Omega_{N-1})$ and using (\ref{Wf1}), we have 
\begin{eqnarray}
X_{n_i}=e^{-\frac{\sqrt{-\omega'} r^2_i}{2}}r_i^{\alpha'_i+\frac{1}{2}}L^{\alpha'_i}_{n_i}(\sqrt{-\omega'} r^2_i),\quad i=1,2.
\end{eqnarray}
$X_{n_i}y_i(\Omega_{N-1})$ are eigenfunctions of $\tilde{H_i}$ with eigenvalues
\begin{eqnarray}
\tilde{\varepsilon_i}=\left[n_i+\frac{1}{2}\alpha'_i+\frac{1}{2}\right]\sqrt{-\omega'},\quad i=1,2.
\end{eqnarray}
Now we define ladder operators
\begin{eqnarray}
&&\tilde{ D^\pm_i}(\sqrt{-\omega'}, r_i)=4\tilde{H_i}\pm2\sqrt{-\omega'} r_i\partial_{r_i}+2\omega' r^2_i\pm\sqrt{-\omega'},\quad i=1,2.
\end{eqnarray}
The action of these operators on the wave functions $X_{n_i}y_i(\Omega_{N-1})$ gives the recurrences formulas (for $i=1,2$)
\begin{eqnarray}
&&\tilde{ D^+_i}(\sqrt{-\omega'}, r_i)X_{n_i}y_i(\Omega_{N-1})=4(n_i+1)\sqrt{-\omega'}X_{n_i+1}y_i(\Omega_{N-1}), 
\\&&
\tilde{ D^-_i}(\sqrt{-\omega'}, r_i)X_{n_i}y_i(\Omega_{N-1})=4(n_i+\alpha'_i)\sqrt{-\omega'}X_{n_i-1}y_i(\Omega_{N-1}).
\end{eqnarray}
The following diagram indicates how $\tilde{D^\pm_i}, i=1,2$ change the quantum numbers.

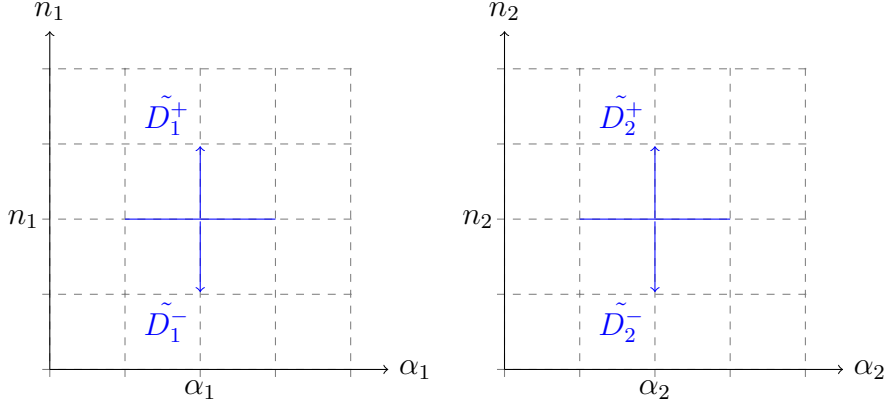
\begin{tikzpicture}
\draw[dashed,color=gray] (-0.1,-0.1) grid (4.1,4.1);

    \draw[->] (0,0) -- (0,4.5) node[above] {$n_1$};
    \draw[ ->] (0,0) -- (4.5,0) node[right] {$\alpha_1$};
       
    \draw[-> ,blue] (2,2.01) -- (2,2.97)node[above left] {$\tilde{D^+_1}$};
    \draw[-> ,blue] (2,1.97) --(2,1.03) node[below left] {$\tilde{D^-_1}$};
    
    \draw[- ,blue] (1,2) -- (3,2);
        \node at (2,0)[below]{$\alpha_1$};
    \node at (0,2)[left]{$n_1$};
      
\end{tikzpicture}
\begin{tikzpicture}
\draw[dashed,color=gray] (-0.1,-0.1) grid (4.1,4.1);

    \draw[->] (0,0) -- (0,4.5) node[above] {$n_2$};
    \draw[ ->] (0,0) -- (4.5,0) node[right] {$\alpha_2$};
       
    \draw[-> ,blue] (2,2.01) -- (2,2.97)node[above left] {$\tilde{D^+_2}$};
    \draw[-> ,blue] (2,1.97) --(2,1.03) node[below left] {$\tilde{D^-_2}$};
    
    \draw[- ,blue] (1,2) -- (3,2);
        \node at (2,0)[below]{$\alpha_2$};
    \node at (0,2)[left]{$n_2$};
      
\end{tikzpicture}

Let us consider the higher order operators $\tilde{L_1}=\tilde{D^+_1}\tilde{ D^-_2}$, $\tilde{L_2}=\tilde{D^-_1}\tilde{ D^+_2}$, $\tilde{H}=\tilde{H_1}+\tilde{H_2}$ and $\tilde{B}=\tilde{H_1}-\tilde{H_2}$. Then the action of these operators provide us the following higher order integrals of motion for the gauge rotated Hamiltonian $\tilde{H}$ 
\begin{eqnarray}
&&\tilde{L_1}Z=-16\omega'(n_1+1)(n_2+\alpha'_2)X_{n_1+1}X_{n_2-1}y_1(\Omega_{N-1})y_2(\Omega_{N-1}) ,
\\&&
\tilde{L_2}Z=-16\omega'(n_2+1)(n_1+\alpha'_1)X_{n_1-1}X_{n_2+1}y_1(\Omega_{N-1})y_2(\Omega_{N-1}),
\\&&
\tilde{L_1}\tilde{L_2} Z=256\omega'^2 n_1(n_2+1)(n_1+\alpha'_1)(n_2+\alpha'_2+1)Z,
\\&&
\tilde{L_2}\tilde{L_1} Z=256\omega'^2 n_2(n_1+1)(n_2+\alpha'_2)(n_1+\alpha'_1+1)Z
\end{eqnarray}
and the cubic algebra $C(3)$ in operator form
\begin{eqnarray}
&[\tilde{L_1},\tilde{H}]=0=[\tilde{L_2},\tilde{H}],&\label{kp5}
\\&[\tilde{L_1},\tilde{B}]=-2\sqrt{-\omega'}\tilde{L_1} ,\qquad [\tilde{L_2},\tilde{B}]=2\sqrt{-\omega'}\tilde{L_2},&\label{kp6}
\end{eqnarray}
\begin{eqnarray}
&\tilde{L_1}\tilde{L_2}& =16\left[(\tilde{B}+\tilde{H}-\sqrt{-\omega'})^2+\frac{\omega'}{\hbar^2}\left\{J^2+4c'_1\hbar^2+\frac{(N-3)^2}{4}\hbar^2\right\}\right]\nonumber\\&&\quad\times\left[(\tilde{B}-\tilde{H}-\sqrt{-\omega'})^2+\frac{\omega'}{\hbar^2}\left\{J^2+4c'_2\hbar^2+\frac{(N-3)^2}{4}\hbar^2\right\}\right],
\nonumber\\&&\label{kp9}
\end{eqnarray}
\begin{eqnarray}
&\tilde{L_2}\tilde{L_1}& =16\left[(\tilde{B}+\tilde{H}+\sqrt{-\omega'})^2+\frac{\omega'}{\hbar^2}\left\{J^2+4c'_1\hbar^2+\frac{(N-3)^2}{4}\hbar^2\right\}\right]\nonumber\\&&\quad\times\left[(\tilde{B}-\tilde{H}+\sqrt{-\omega'})^2+\frac{\omega'}{\hbar^2}\left\{J^2+4c'_2\hbar^2+\frac{(N-3)^2}{4}\hbar^2\right\}\right].\nonumber\\&&\label{kp10}
\end{eqnarray}
Applying the simple gauge rotations to $D'^{\pm}_i$, we can obtain the corresponding integrals of motion and cubic algebra of Hamiltonian $H'$.  

The first order integrals of motion $L_{ij}$ generate $so(N-1)$ Lie algebra
\begin{eqnarray}
&[L_{ij},L_{kl}]&= i(\delta_{ik}L_{jl}+ \delta_{jl}L_{ik}-\delta_{il}L_{jk}-\delta_{jk}L_{il})\hbar, 
\end{eqnarray}
where $i, j, k, l=1, ..., N-1.$ Moreover,
\begin{eqnarray}
[L_{ij},\tilde{L_1}]=0=[L_{ij},\tilde{L_2}],\quad[L_{ij},\tilde{H}]=0=[L_{ij},\tilde{B}].
\end{eqnarray}
Thus the full symmetry algebra is a direct sum of the cubic algebra $C(3)$ and $so(N-1)$ Lie algebra (i.e., $C(3)\oplus so(N-1))$.

\subsection{Unirreps and Energy spectrum}

The cubic algebra (\ref{kp5})-(\ref{kp10}) has already the form of the deformed oscillator algebra with $\aleph=\frac{\tilde{B}}{2\gamma}$, $b^{\dagger}=\tilde{L_1}$ and $b=\tilde{L_2}$. The structure function is
\begin{eqnarray}
\Phi(x;u,\tilde{H})&=&16\left[2(x+u)\gamma+\tilde{H}-(1-\alpha_1)\gamma\right]\left[2(x+u)\gamma+\tilde{H}-(1+\alpha_1)\gamma\right]\nonumber\\&&\times\left[2(x+u)\gamma-\tilde{H}-(1-\alpha_2)\gamma\right]\left[2(x+u)\gamma-\tilde{H}-(1+\alpha_2)\gamma\right],\nonumber\\&&
\end{eqnarray}
where $\gamma=\sqrt{-\omega'}$ and $u$ is an arbitrary constant to be determined. We should impose the following constraints on the structure function in order for the representations to be finite dimensional
\begin{equation}
\Phi(p+1; u,\tilde{\varepsilon})=0,\quad \Phi(0;u,\tilde{\varepsilon})=0,\quad \Phi(x)>0,\quad \forall x>0,\label{pro2}
\end{equation}
where $p$ is a positive integer. The solutions of the constraints give the energy $\tilde{\varepsilon}$, the arbitrary constant $u$ and the structure function (i.e. the $(p+1)$-dimensional unirreps )
\begin{eqnarray}
u=\frac{-\tilde{\varepsilon}+\gamma(1+\alpha'_1)}{2\gamma},\quad\tilde{\varepsilon}=\frac{\gamma}{2}(2+2p+k_1 \alpha'_1+k_2 \alpha'_2),
\end{eqnarray}
\begin{eqnarray}
\Phi(x) &=&16x\gamma^4(\alpha'_1+x)\left[2+2p-2x-(1-k_1)\alpha'_1+(1+k_2)\alpha'_2\right]\nonumber\\&&\times\left[2+2p-(1-k_1)\alpha'_1-(1-k_2)\alpha'_2\right],
\end{eqnarray}
where $k_1=\pm 1$, $k_2=\pm 1$.
By equivalence of spectra between Hamiltonians related by the gauge transformations, we have 
\begin{eqnarray}
\varepsilon'=\frac{\gamma}{2}(2+2p+k_1 \alpha'_1+k_2 \alpha'_2).\label{Eng}
\end{eqnarray}
The coupling constant metamorphosis provides $\varepsilon'\leftrightarrow-\beta_0$ and $\varepsilon\leftrightarrow\omega'$ and give a correspondence between $\gamma=\sqrt{-\omega'}$ and the energy $\varepsilon$ of the original Hamiltonian $H$. From (\ref{Eng}) we obtain 
\begin{eqnarray}
E=\frac{-c^2_0}{\hbar^2\left(p+1+\frac{k_1\alpha'_1+k_2\alpha'_2}{2}\right)^2}.\label{en4}
\end{eqnarray}
Making the identifications $p=n_1+n_2$, $k_1= 1$, $k_2= 1$, then (\ref{en4}) coincides with the physical spectra (\ref{en2}).

We remark that the recurrence method gives 4th-order integrals of motion and higher rank cubic algebra
for the Stackel equivalent system. This algebra has already a factorized form and much easier to handle than the quadratic algebra $Q(3)$ obtained in \cite{FH1} by using ansatz and the direct approach. 

\section{Conclusion}
The main result of this paper is extending the recurrence approach, that is a constructive approach, to the $N$-dimensional superintegrable Kepler-Coulomb system with non-central terms and the double singular oscillators of type $(n,N-n)$ to obtain algebraic derivations of their spectra. For both cases, we obtained 4th-order integrals of motion and corresponding higher rank cubic algebras. One interesting feature of these algebras is the fact they admit a factorized form which enable to simplify the calculation of the realizations in terms of deformed oscillator algebras and the corresponding unirreps. 

Let us point out that the classical analogs of the recurrence relations \cite{Mar4} have been developed and exploited to generate superintegrable systems with higher order integrals and polynomial Poisson algebras \cite{Mar7,Mar8}. Recently another classical method based on subgroup coordinates \cite{Mil2, Mil3, Des1} has been introduced. One interesting open problem, that we study in a future paper, would be to extend these approaches for the classical analog of the two models considered in this paper. Moreover, the application of the co-algebra approach \cite{Bal1,Rig1, Pos1} to these new models also remains to be investigated.

\end{document}